\documentclass[journal]{IEEEtran}
\usepackage[dvips]{graphicx}
\usepackage{verbatim}	
\usepackage[cmex10]{amsmath}
\usepackage{amssymb}
\usepackage{ifthen}
\usepackage{pgf}
\usepackage{tikz}
\usepackage{epsfig}

\def\paira#1{{\langle{#1}\rangle}} 
\def\ddd#1{{\downarrow\!\!{#1}}}
\def\uuu#1{{\uparrow\!\!{#1}}}

\def\Size{\mathop{\rm Size}\nolimits}              

\def\nauty/{{\tt nauty}}
\def\magma/{{\tt magma}}
\def\Magma/{{\tt Magma}}

\newtheorem{theorem}{Theorem}
\newtheorem{lemma}[theorem]{Lemma}

\newtheorem{corollary}[theorem]{Corollary}

\newtheorem{definition}[theorem]{Definition}

\def\eqref#1{(\ref{#1})}    			

\def\de{\delta}
\def\ga{p}  
\newcommand{\ep}{\varepsilon}            
\newcommand{\epsi}{\varepsilon}            
\def\epsimax{{\epsi_{\max}}}

\def\si{\sigma}
\def\ze{\zeta}

\def\cC{{\cal C}}

\def\cG{{\cal G}}
\def\cH{{\cal H}}

\def\cP{{\cal P}}
\def\cQ{{\cal Q}}
\def\cR{{\cal R}}
\def\cS{{\cal S}}

\def\cV{{\cal V}}

\def\sC{{\hbox{\sf\slshape C}}}

\def\sE{{\hbox{\sf\slshape E}}}

\def\sP{{\hbox{\sf\slshape P}}}
\def\sS{{\hbox{\sf\slshape S}}}

\def\cc{\mathbf{c}}
\def\ee{\mathbf{e}}
\def\gg{\mathbf{e}}
\def\vv{\mathbf{v}}

\def\xx{\mathbf{x}}
\def\yy{\mathbf{y}}

\def\Fnum{{\mathbb{F}}} 
\def\Rnum{{\mathbb{R}}} 

\def\zero{\mathbf{0}}
\def\prob{\mathop{\rm Prob}\nolimits}              

\def\sumno{\sum\nolimits}

\newdimen\eqnarraycolsep
\makeatletter
\def\eqnarray{\stepcounter{equation}\let\@currentlabel=\theequation
\global\@eqnswtrue
\global\@eqcnt\z@\tabskip\@centering\let\\=\@eqncr
$$\halign to \displaywidth\bgroup\@eqnsel\hskip\@centering
  $\displaystyle\tabskip\z@{##}$&\global\@eqcnt\@ne
  \hskip 2\eqnarraycolsep \hfil${##}$\hfil
  &\global\@eqcnt\tw@ \hskip 2\eqnarraycolsep
  $\displaystyle\tabskip\z@{##}$\hfil
   \tabskip\@centering&\llap{##}\tabskip\z@\cr}

\def\thebibliography#1{\section*{References\@mkboth
 {REFERENCES}{REFERENCES}}\typeout{references:}
 \list
 {[\arabic{enumi}]}{\settowidth\labelwidth{[#1]}\leftmargin\labelwidth
 \advance\leftmargin\labelsep
 \usecounter{enumi}}
 \def\newblock{\hskip .11em plus .33em minus .07em}
 \sloppy\clubpenalty4000\widowpenalty4000
 \sfcode`\.=1000\relax}

\makeatother

{\obeyspaces\gdef {\hglue.5em\relax}}

{\catcode`\^^M=\active %
}

{\catcode`\=0 \catcode`\\=12 catcode`^^M=active %
 gdefthatisit^^M#1\endliteral{#1endgroup%
   noindentignorespaces}}

\begin{document}

\title{Optimal hash functions for approximate matches on the $n$-cube}

\author{Daniel~M.~Gordon, Victor~S.~Miller and Peter~Ostapenko
\thanks{D. Gordon and P. Ostapenko are with the IDA Center for
  Communications Research, 4320 Westerra Court, San Diego, 92121}%
\thanks{V. Miller is with the IDA Center for Communications Research, 805 Bunn Drive,
Princeton, New Jersey 08540}}

\maketitle

\begin{abstract}



One way to find near-matches in large datasets is to use hash functions
\cite{broder}, \cite{kwz}.  In recent years locality-sensitive
hash functions for various metrics have been given; for the Hamming metric
projecting onto $k$ bits is simple hash function that
performs well.

In this paper we investigate alternatives to projection.
For various parameters hash functions given
by complete decoding algorithms for error-correcting codes work better, and
asymptotically random codes perform better than projection.

\end{abstract}

\maketitle



\section{Introduction}
\label{sec:intro}

Given a set of $M$ $n$-bit vectors, a classical problem is to 
quickly identify ones which are close in Hamming distance.
This problem has applications
in numerous areas, such as information retrieval and DNA sequence
comparison.
The nearest-neighbor problem is to find a vector close to a given one,
while the closest-pair problem is to find the pair in the set with the smallest
Hamming distance.  Approximate versions of these problems allow an
answer where the distance may be a factor of $(1+\varepsilon)$ larger than the best possible.

One approach (\cite{broder}, \cite{gim}, \cite{kwz}) is {\em
locality-sensitive hashing} (LSH).
A
family of hash functions $\cH$ is called $(r,cr,p_1,p_2)$-sensitive if
for any two points $\xx$, $\yy \in \cV$,
\begin{itemize}
\item if $d(\xx,\yy) \leq r$, then $\prob(h(\xx) = h(\yy)) \geq p_1$,
\item if $d(\xx,\yy) \geq cr$, then $\prob(h(\xx) = h(\yy)) \leq p_2$.
\end{itemize}

Let $\rho = \log(1/p_1)/\log(1/p_2)$.  
An LSH scheme can be used to solve the approximate nearest neighbor
problem for $M$ points in time $O(M^\rho)$.  Indyk and Motwani
\cite{im} showed that projection has $\rho = 1/c$.

The standard hash to use is
projection onto $k$ of the $n$ coordinates \cite{gim}.
An alternative family of hashes is based on minimum-weight decoding
with error-correcting codes
\cite{be}, \cite{w}.
A $[n,k]$ code $\cC$ with a complete
decoding algorithm defines a hash $h^\cC$, where each
$\vv \in \cV :=\Fnum_2^n$ is mapped to the codeword $\cc \in \cC \subset \cV$
to which $\vv$ decodes.
Using linear codes for hashing schemes has been independently suggested many
times; see \cite{be}, \cite{dhlnp}, and the
patents~\cite{patent2} and \cite{w}.

In \cite{be} the binary Golay code was suggested to find approximate
matches in bit-vectors.  Data is provided that suggests it is effective,
but it is still not clear when the Golay or other codes work better than
projection.  In this paper we attempt to quantify this, using tools
from coding theory.

Our model is somewhat different from the usual LSH literature.
We are interested in the scenario where we have
collection of $M$ random points of $\cV$, one of which, $\xx$, has been
duplicated with errors.  The error vector $\ee$ has each bit nonzero
with probability $\ga$.
Let
$\sP^\cC(p)$ be the probability that $h^\cC(\xx)=h^\cC(\xx+\ee)$.
Then the probability of collision of two points $\xx$ and $\yy$ is
\begin{itemize}
\item if $\yy=\xx+\ee$, then $\prob(h(\xx) = h(\yy)) = \tilde{p}_1 = \sP^\cC(p)$,
\item if $\yy \neq \xx+\ee$, then $\prob(h(\xx) = h(\yy)) = \tilde{p}_2 = 2^{-k}$.
\end{itemize}

Then the number of elements that hash to $h(\xx)$ will be about
$M/2^k$, and the probability that one of these will be $\yy=\xx+\ee$ is 
$\sP^\cC(p)$.  If this fails, we may try again with a new hash, say
the same one applied after shifting the $M$ points by a fixed vector, and
continue until $\yy$ is found.

Let $\rho=\log(1/\tilde{p}_1)/\log(1/\tilde{p}_2)$ as for LSH.
Taking $2^k \approx M$, we expect to find $\yy$ in time
$$
\frac{M}{2^k \sP^\cC(p)} = O(M^\rho).
$$
As with LSH, we want to optimize this by minimizing $\rho$,
i.e. finding a hash function minimizing $\sP^\cC(p)$.

For a linear code with a complete translation-invariant decoding algorithm
(so that $h(\xx)=\cc$ implies that $h(\xx+\cc')=\cc+\cc'$),
studying $\sP^{\cC}$ is equivalent to studying the properties
of the set $\cS$ of all points in $\cV$ that decode to ${\bf 0}$.
In Section~\ref{sec:sets} and
the appendix we systematically investigate sets of size $\leq 64$.

Suppose that we pick a random $\xx \in \cS$.
Then the probability that $\yy = \xx + \ee$ is in $\cS$ is
\begin{equation}
  \label{eq:pue}
\sP_\cS(\ga) =  \frac{1}{|\cS|}  \sum_{\xx,\yy \in \cS} \ga^{d(\xx,\yy)}
  (1-\ga)^{n-d(\xx,\yy)}.
\end{equation}

This function has been studied extensively in the setting of
error-detecting codes \cite{kk}.  In that literature, $\cS$ is a code,
$\sP_\cS(\ga)$ is the probability of an undetected error, and the goal
is to minimize this probability.  Here, on the other hand, we will
call a set {\em optimal} for $\ga$ if no set in $\cV$ of size $|\cS|$
has greater probability.

As the error rate $\ga$ approaches $1/2$, this coincides
with the definition of {\em distance-sum optimal sets}, which were
first studied by Ahlswede and Katona \cite{ak}.

The error exponent
of a code $\sC$ is
$$
\sE^\cC(\ga) =
-\frac{1}{n}  \lg \sP^\cC(\ga).
$$
In this paper $\lg$ denotes log to base 2.  We are interested in
properties of the error exponent over
codes of rate $R=k/n$ as $n \to \infty$.
Note that $\rho = E^\cC(p)/R$, so minimizing the error exponent will
give us the best code to use for finding closest pairs.
In Section~\ref{sec:shannon} we will show that
hash functions from random (nonlinear)
codes have a better error exponent than projection.

\section{Hash Functions From Codes}
\label{sec:codes}

For a set $\cS \subset \cV$, let
$$
A_i = \# \{ (\xx,\yy) : \xx,\yy \in \cS \hbox{ and } d(\xx,\yy) = i \}
$$
count the number of pairs of words in $\cS$ at distance $i$.
The {\em distance distribution function} is
\begin{equation}
  \label{eq:gf}
A(\cS,\ze) := \sum_{i=0}^n A_i \ze^i .
\end{equation}

This function is directly connected to $\sP_\cS(\ga)$ \cite{kk}.
If $\xx$ is a random element of $\cS$, and $\yy = \xx +e$, where $e$
is an error vector where each bit is nonzero with probability $\ga$,
then the
probability that $\yy \in \cS$ is
\begin{eqnarray} \label{eq:pr_ue}
 \sP_\cS(\ga) & := &
\frac{1}{|\cS|}
\sum_{\xx,\yy \in \cS} \ga^{d(\xx,\yy)} (1-\ga)^{n-d(\xx,\yy)} \\ \nonumber
& = & \frac{1}{|\cS|} \sum_{i=0}^n A_i \ga^i (1-\ga)^{n-i} \\ \nonumber
& = & \frac{(1-\ga)^n}{|\cS|} A\left(\cS ,\frac{\ga}{1-\ga} \right).
\end{eqnarray}

In this section we will evaluate (\ref{eq:pr_ue}) for projection and
for perfect codes, and then consider other linear codes.

\subsection{Projection} 
\label{subsec:projection}

The simplest hash is to project vectors in $\cV$
onto $k$ coordinates.
Let \emph{$k$-projection} denote the $[n,k]$ code $\cP_{n,k}$
corresponding to
this hash.
The associated $\cS$ of vectors mapped to $\zero$
is an $2^{n-k}$-subcube of $\cV$.
The distance distribution function is
\begin{equation}
\label{eq:proj_ddf}
A(\cS,\ze) = (2(1+\ze))^{n-k} \, ,
\end{equation}
so the probability of collision is
\begin{equation}
  \label{eq:proj_prob}
 \sP^{\cP_{n,k}}(\ga)
 = \frac{(1-\ga)^n}{2^{n-k}} \left(\frac{2}{1-\ga}\right)^{n-k}
 = (1-\ga)^k.
\end{equation}

$\cP_{n,k}$ is not a good error-correcting code, but
for sufficiently small error probability its hash function is optimal.

\begin{theorem}
\label{thm:gamma_small}
Let $\cS$ be the $2^{n-k}$-subcube of
$\cV$.
For any error probability $\ga \in (0, \, 2^{-2(n-k)})$,
$\cS$ is an optimal set,
and so $k$-projection is an optimal hash.
\end{theorem}

\begin{proof}
The distance distribution function for $\cS$ is
$$
A(\cS,\ze) = 2^{n-k}(1+\ze)^{n-k}.
$$

The edge isoperimetric inequality for an $n$-cube \cite{harper2} states that
\begin{lemma} \label{lem:iso}
Any subset $S$ of the vertices of the $n$-dim\-en\-sion\-al cube
$Q_n$ has at most
$$
\frac{1}{2} |S| \lg |S|
$$
edges between vertices in $S$, with equality if and only if $S$ is a
subcube.
\end{lemma}

Any set $\cS'$ with $2^{n-k}$ points
has distance distribution function
$$A(\cS',\ze) = \sum_{i=0}^k c_i \ze^i,$$
where $c_0 = 2^{n-k}$, $c_1 < (n-k) 2^{n-k}$ by Lemma~\ref{lem:iso},
and the sum of the $c_i$'s is $2^{2(n-k)}$.
By (\ref{eq:proj_prob}) the probability of collision is
$(1-\ga)^n 2^{n-k}A(\sS',\ga/(1-\ga))$.
\begin{eqnarray*}
A(\cS',\ze)
& \leq  & 2^{n-k} + \ze((n-k)2^{n-k}-1)\\
&& + \ze^2
 \left(2^{2(n-k)}-({n-k}+1)2^{n-k}+1\right),
\end{eqnarray*}
and
\begin{eqnarray*}
 \lefteqn{A(\cS,\ze) - A(\cS',\ze)} \quad \\
  &\geq& \ze - \ze^2 \left(2^{2(n-k)}+
   2^{{n-k}-1}\left({n-k}^2+{n-k}+2\right)+1 \right) \\
  &>& \ze-\ze^2(2^{2(n-k)}-1).
\end{eqnarray*}
This is positive if $\ga<1/2$ and $(1-\ga)/\ga > 2^{2(n-k)}-1$, i.e.,
for $\ga<2^{-2(n-k)}$.
\end{proof}

\subsection{Concatenated Hashes} 
\label{subsec:concat}

Here we show that if $h$ and $h'$ are good hashes,
then the concatenation is as well.
First we identify $\cC$ with $\Fnum_2^k$
and treat $h^\cC$ as a hash $h$ from $\Fnum_2^n \to \Fnum_2^k$.
We denote $\sP^\cC$ by $\sP^h$.
From $h : \Fnum_2^n \to \Fnum_2^k$
and $h' : \Fnum_2^{n'} \to \Fnum_2^{k'}$,
we get a concatenated hash $(h,h') : \Fnum_2^{n+n'} \to \Fnum_2^{k+k'}$.

\begin{lemma}
Fix $\ga \in (0,1/2)$.
Let $h$ and $h'$ be hashes.  Then
$$
 \min \{ \sE^h(\ga), \sE^{h'}(\ga) \}
 \leq \sE^{(h,h')(\ga)}
 \leq \max \{ \sE^{h}(\ga), \sE^{h'}(\ga) \} \, ,
$$
with strict inequalities if $\sE^{h}(\ga) \not= \sE^{h'}(\ga)$.
\end{lemma}

\begin{proof}
Since $\ga$ is fixed, we drop it from the notation.
Suppose  $\sE^h \leq \sE^{h'}$.  Then
$$
\frac{\lg \sP^h}{n}
 \leq \frac{\lg \sP^h + \lg \sP^{h'}}{n+n'}
 \leq \frac{\lg \sP^{h'}}{n'}.
$$
Since $\sP^{(h,h')} = \sP^h \, \sP^{h'}$, we have
$\sE^h \leq \sE^{(h,h')} \leq \sE^{h'}$.
\end{proof}

\subsection{Perfect Codes} 
\label{subsec:perfect}

An {\em $e$-sphere} around a vector $\xx$ is the set of all vectors
$\yy$ with $d(\xx,\yy) \leq e$.
An $[n,k,2e+1]$ code
is {\em perfect} if
the $e$-spheres around codewords cover $\cV$.  Minimum weight decoding with
perfect codes is a
reasonable starting point for hashing schemes, since all vectors are
closest to a unique codeword.  The only perfect binary codes are
trivial repetition codes, the Hamming codes, and the binary Golay
code.  Repetition codes do badly, but the other perfect codes give
good hash functions.

\subsubsection{Binary Golay Code} 
\label{subsec:golay}

The $[23,12,7]$ binary Golay code $\cG$ is an important perfect code.  The
3-spheres around each code codeword cover $\Fnum_2^{23}$.
The 3-sphere around $\zero$
in the 23-cube has distance distribution function
\begin{eqnarray*}
\lefteqn{2048 + 11684 \ze + 128524 \ze^2 + 226688 \ze^3}
  \qquad\qquad \\
 && \hbox{}
 + 1133440 \ze^4 + 672980 \ze^5 + 2018940 \ze^6 \, .
\end{eqnarray*}
From this we find $\sE^\cG(\ga)>\sE^{\cP_{23,12}}(\ga)$
for $\ga \in (0.2555, \, 1/2)$.

\subsubsection{Hamming Codes} 
\label{subsec:hamming}

Aside from the repetition codes and the Golay code,
the only perfect binary codes are the Hamming codes.
The $[2^m-1,\, 2^m-m-1,\, 3]$ Hamming code $\cH_m$
corrects one error.

The distance distribution function for a 1-sphere is
\begin{equation}
\label{eq:hamming_ddf}
2^m + 2(2^m-1) \ze + (2^m-1)(2^m-2) \ze^2,
\end{equation}
so the probability of collision $\sP^{\cH_m}(\ga)$ is
\begin{eqnarray}
 \label{eq:ham_prob}
 \frac{(1-\ga)^{2^m-1}}{2^m}(2^m &+& 2(2^m-1)
 \frac{\ga}{1-\ga}\\
&+& (2^m-1)
 (2^m-2) \frac{\ga^2}{(1-\ga)^2} ) \nonumber
\end{eqnarray}

Table~\ref{tab:ham} gives the crossover
error probabilities where the first few Hamming codes become better than projection.

\begin{table}[!t]
  \caption{Crossover error probabilities $\ga$ for Hamming codes $\cH_m$.}
  \label{tab:ham}
  \centering
  \begin{tabular}{|cc|c|}
\hline
$m$ & $k$ & $\ga$ \\  \hline
4   & 11 & $0.2826$ \\
5   & 26 & $0.1518$ \\
6   & 57 & $0.0838$ \\
7   &120 & $0.0468$ \\ \hline
  \end{tabular}
\end{table}


\begin{theorem}
  \label{thm:hamming_wins}
For any $m>4$ and $p>m/(2^m-m)$, the Hamming code $\cH_m$ beats
$(2^m-m-1)$-projection.
\end{theorem}

\begin{proof}
  The difference between the distribution functions of the cube and
  the 1-sphere in dimension $2^m-1$ is
\begin{eqnarray}
\label{eq:fm}
    f_m(\ze) &:=& A(\cS,\ze) - A(\cH_m,\ze) \\
&=& 2^m(1+\ze)^m \nonumber \\
&& - (2^m + 2(2^m-1) \ze + (2^m-1)(2^m-2) \ze^2).\nonumber
  \end{eqnarray}
  We will show that, for $m \ge 4$, $f_m(\ze)$ has exactly one root in $(0,1)$, denoted
  by $\alpha_m$, and that $\alpha_m \in \left( (m-2)/2^m,m/2^m\right)$.

  We calculate
\begin{eqnarray*}
    f_m(\ze) &=& ((m-2)2^m+1)\ze \\
&&-\left(2^{2m} -\left(3+\binom{m}{2}\right)2^m +
      2\right)\ze^2 \\
&&+ 2^m\sum_{i=3}^m \binom{m}{i} \ze^i.
\end{eqnarray*}
  All the coefficients of $f_m(\ze)$ are non-negative with the exception
  of the coefficient of $\ze^2$, which is negative for $m \ge 2$.  Thus,
  by Descartes' rule of signs $f(\ze)$ has 0 or 2 positive roots.
  However, it has a root at $\ze=1$.  Call the other positive root
  $\alpha_m$.
  We have $f_m(0) = f_m(1) = 0$, and since $f'(0) = (m-2) 2^m+2 >0$
  and $f'(1) = 2^{2m-1}(m-4)+2^{m+2}-2>0$ for $m\geq 4$, we must have
  $\alpha_m<1$ for $m\geq 4$.

For $p>\alpha_m$ the Hamming code $\cH_m$ beats
projection.  
  
Using (\ref{eq:fm}) and Bernoulli's inequality, it is easy to show
that $f_m(\ze)>0$ for $\ze < c(m-2)/2^m$ for any $c<1$ and $m \geq 4$.
For the other direction, we may use Taylor's theorem to show
$$
2^m\left(1+\frac{m}{2^m}\right)^m < 2^m + m^2 + \frac{m^4}{2^{m+1}} 
\left(1+\frac{m}{2^m}\right)^{m-2}.
$$
Plugging this into (\ref{eq:fm}), we have that $f_m(m/2^m)<0$ for $m>6$.

\end{proof}


\subsection{Other Linear Codes} 
\label{subsec:linear}

The above codes give hashing strategies for a few values of $n$ and
$k$, but we would like hashes for a wider range.
For a hashing
strategy using error-correcting codes, we need a code with an
efficient {\em complete decoding} algorithm; that is a way to map
every vector to a codeword.  Given a translation invariant decoder, we
may determine $\cS$, the set of vectors that decode to ${\bf 0}$,
in order to compare strategies as the error probability changes.

\begin{figure}[!t]
\centering
    \leavevmode
    \begin{tikzpicture}[xscale=0.23,yscale=9]
      \draw (0,0) -- (30,0);
      \draw (30,0) -- (30,0.5);
      \draw (0,0) -- (0,0.5);
      \draw (0,0.5) -- (30,0.5);

      \draw (14,-0.07) node [left] {$k$};
      \draw (-5,0.25) node [below] {$\ga$};

        \path[draw=black]
            (23,0.45) ellipse (5pt and 0.1pt) --
            (25,0.45) ellipse (5pt and 0.1pt) node [right]
            {\footnotesize $d=3$};

        \path[draw=black]
           (8 ,0.456) ellipse (5pt and 0.1pt) node (3_8) [] {}
        -- (9 ,0.393) ellipse (5pt and 0.1pt) node (3_9) [] {}
	--  (10 ,0.333) ellipse (5pt and 0.1pt) node (3_10) [] {}
	--  (11 ,0.2826) ellipse (5pt and 0.1pt) node (3_11) [left] {\footnotesize $\cH_4$}
	--  (12 ,0.28656) ellipse (5pt and 0.1pt) node (3_12) [] {}
	--  (13 ,0.28731) ellipse (5pt and 0.1pt) node (3_13) [] {}
	--  (14 ,0.28503) ellipse (5pt and 0.1pt) node (3_14) [] {}
	--  (15 ,0.279944) ellipse (5pt and 0.1pt) node (3_15) [] {}
	--  (16 ,0.272384) ellipse (5pt and 0.1pt) node (3_16) [] {}
	--  (17 ,0.262737) ellipse (5pt and 0.1pt) node (3_17) [] {}
	--  (18 ,0.25145) ellipse (5pt and 0.1pt) node (3_18) [] {}
	--  (19 ,0.239) ellipse (5pt and 0.1pt) node (3_19) [] {}
	--  (20 ,0.2259) ellipse (5pt and 0.1pt) node (3_20) [] {}
	--  (21 ,0.2125) ellipse (5pt and 0.1pt) node (3_21) [] {}
	--  (22 ,0.1992) ellipse (5pt and 0.1pt) node (3_22) [] {}
	--  (23 ,0.1864) ellipse (5pt and 0.1pt) node (3_23) [] {}
	--  (24 ,0.1741) ellipse (5pt and 0.1pt) node (3_24) [] {}
	--  (25 ,0.1626) ellipse (5pt and 0.1pt) node (3_25) [] {}
	-- (26 ,0.1518) ellipse (5pt and 0.1pt) node (3_26) [below] {\footnotesize $\cH_5$}
 	--  (27 ,0.1529) ellipse (5pt and 0.1pt) node (3_27) [] {}
 	--  (28,0.1535) ellipse (5pt and 0.1pt) node (3_28) [] {}
 	--  (29,0.1538) ellipse (5pt and 0.1pt) node (3_29) [] {}
 	--  (30,0.1537) ellipse (5pt and 0.1pt) node (3_30) [] {};

        \path[draw=red]
            (23,0.42) ellipse (5pt and 0.1pt) --
            (25,0.42) ellipse (5pt and 0.1pt) node [right]
            {\footnotesize $d=5$};
        \path[draw=red]
	    (9 ,0.453) ellipse (5pt and 0.1pt) node (5_9) [] {}
	--  (10 ,0.45346) ellipse (5pt and 0.1pt) node (5_10) [] {}
	--  (11 ,0.424057) ellipse (5pt and 0.1pt) node (5_11) [] {}
	--  (12 ,0.399166) ellipse (5pt and 0.1pt) node (5_12) [] {}
	--  (13 ,0.37273) ellipse (5pt and 0.1pt) node (5_13) [] {}
	--  (14 ,0.34225) ellipse (5pt and 0.1pt) node (5_14) [] {}
	--  (15 ,0.355389) ellipse (5pt and 0.1pt) node (5_15) [] {}
	--  (16 ,0.335497) ellipse (5pt and 0.1pt) node (5_16) [] {}
	--  (17 ,0.318722) ellipse (5pt and 0.1pt) node (5_17) [] {}
	--  (18 ,0.302035) ellipse (5pt and 0.1pt) node (5_18) [] {}
	--  (19 ,0.28666) ellipse (5pt and 0.1pt) node (5_19) [] {}
	--  (20 ,0.273291) ellipse (5pt and 0.1pt) node (5_20) [] {}
	--  (21 ,0.260697) ellipse (5pt and 0.1pt) node (5_21) [] {} 
	--  (22 ,0.248295) ellipse (5pt and 0.1pt) node (5_22) [] {}
	--  (23 ,0.236963) ellipse (5pt and 0.1pt) node (5_23) [] {}
	--  (24 ,0.2611) ellipse (5pt and 0.1pt) node (5_24) [] {}
	--  (25 ,0.2542) ellipse (5pt and 0.1pt) node (5_25) [] {};

        \path[draw=green]
            (23,0.39) ellipse (5pt and 0.1pt) --
            (25,0.39) ellipse (5pt and 0.1pt) node [right]
            {\footnotesize $d=7$};

        \path[draw=green]
	    (8 ,0.48534) ellipse (5pt and 0.1pt) node (7_8) [] {}
	--  (9 ,0.42879) ellipse (5pt and 0.1pt) node (7_9) [] {}
	--  (10 ,0.372251) ellipse (5pt and 0.1pt) node (7_10) [] {}
	--  (11 ,0.315547) ellipse (5pt and 0.1pt) node (7_11) [] {}
	-- (12 ,0.2555) ellipse (5pt and 0.1pt) node (7_12) [below]
	    {\footnotesize $\cG$}
	--  (13 ,0.34646) ellipse (5pt and 0.1pt) node (7_13) [] {}
	--  (14 ,0.32713) ellipse (5pt and 0.1pt) node (7_14) [] {}
	--  (15 ,0.32515) ellipse (5pt and 0.1pt) node (7_15) [] {}
	--  (16 ,0.309727) ellipse (5pt and 0.1pt) node (7_16) [] {}
	--  (17 ,0.29765) ellipse (5pt and 0.1pt) node (7_17) [] {}
	--  (18 ,0.29002) ellipse (5pt and 0.1pt) node (7_18) [] {}
	--  (19 ,0.279135) ellipse (5pt and 0.1pt) node (7_19) [] {}
	--  (20 ,0.2675) ellipse (5pt and 0.1pt) node (7_20) [] {}
	--  (21 ,0.2577) ellipse (5pt and 0.1pt) node (7_21) [] {};

  \foreach \x/\xtext in {0/0, 5/5, 10/10, 15/15, 20/20, 25/25, 30/30}
    \draw[shift={(\x,0)},gray,very thin] (0,0.5) -- (0,0) node[below] {\footnotesize $\xtext$};


  \foreach \y/\ytext in {0/0,0.05/0.05,0.1/0.1,0.15/0.15,0.2/0.2,0.25/0.25,0.3/0.3,0.35/0.35,0.4/0.4,0.45/0.45,0.5/0.5}
    \draw[shift={(0,\y)},gray,very thin] (30,0) -- (0,0) node[left] {\footnotesize $\ytext$};
    \end{tikzpicture}
\caption{Crossover error probabilities for minimum length linear codes.}\label{fig:cross}
\end{figure}
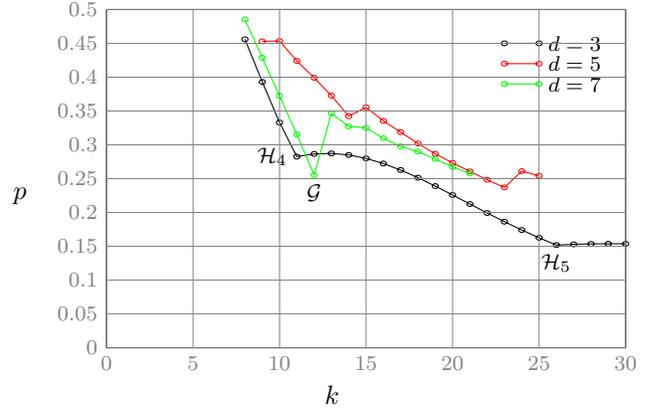

\Magma/~\cite{magma} has a built-in database of linear codes over
$\Fnum_2$ of length up to 256.
Most of these do not come with efficient complete decoding
algorithms, but \magma/ does provide syndrome decoding.
Using this database new hashing schemes were found.
For each dimension $k$ and minimum distance $d$, an $[n,k,d]$ binary
linear code with minimum length $n$ was chosen for testing.\footnote{The
  \magma/ call {\tt BLLC(GF(2),k,d)} was used to choose a code.}
(This criterion excludes any codes formed by concatenating
  with a projection code.)
For each code there is an error probability above which the code beats projection.
Figure~\ref{fig:cross} shows these crossover probabilities.
Not surprisingly, the $[23,12,7]$ Golay code $\cG$ and Hamming codes
$\cH_4$ and $\cH_5$ all do well.  The facts that
concatenating the
Golay code with projection beats the chosen code for
$13 \leq k \leq 17$
and concatenating $\cH_m$ with projection beats the chosen codes
for $27 \leq k \leq 30$
show that factors other than minimum length
are important in determining an optimal hashing code.

As linear codes are subspaces of $\Fnum_2^n$, lattices are
subspaces of $\Rnum^n$.  
The 24-dimensional Leech lattice is closely related to the Golay code, and also has
particularly nice properties.  It was used in \cite{ai} to construct a
good LSH for $\Rnum^n$.

\section{Optimal Sets}
\label{sec:sets}


In the previous section we looked at the performances
of sets associated with various good error-correcting codes.
However, 
the problem of determining optimal sets $\cS \subset \Fnum_2^n$
is of independent interest.



The general question of finding an optimal set of size $2^t$ in
$\cV$ for an error probability $\ga$ is quite hard.
In this section we will find the answer for $t \leq 6$,
and look at what happens when $\ga$ is
near $1/2$.

\subsection{Optimal Sets of Small Size} 
\label{subsec:nauty}

For a vector $\xx = (x_1,\ldots,x_n) \in \cV$, let
$r_i(\xx)$
be $\xx$ with the $i$-th coordinate complemented,
and let
$s_{ij}(\xx)$
be $\xx$ with the $i$-th and $j$-th coordinates switched.

\begin{definition}
Two sets are isomorphic if one can be gotten from the other by
a series of $r_i$ and $s_{ij}$ transformations.
\end{definition}

\begin{lemma}
If $\cS$ and $\cS'$ are isomorphic,
then $\sP_\cS(p) = \sP_{\cS'}(p)$ for all $p \in [0,1]$.
\end{lemma}

The corresponding non-invertible transformation are:
\begin{eqnarray}
\label{eq:rho}
 \rho_i(\xx) &:=& (x_1,x_2,\ldots,x_{i-1},0,x_{i+1},\ldots x_n) \, , \\
\nonumber
 \si_{ij}(\xx) &:=&
   \left\{
   \begin{array}{ll}
     \xx,  & x_{\min(i,j)}=0, \\
\label{eq:si}
     s_{ij}(\xx), & x_{\min(i,j)}=1.
   \end{array}
   \right.
\end{eqnarray}

\begin{definition}
A set $\cS \subset \cV$ is a {\em down-set} if
$\rho_i(\cS) \subset \cS$ for all $i \leq n$.
\end{definition}

\begin{definition}
A set $\cS \subset \cV$ is {\em right-shifted} if
$\si_{ij}(\cS) \subset \cS$ for all $i,j \leq n$.
\end{definition}

\begin{theorem}\label{thm:rightdown}
If a set $\cS$ is optimal, then it is isomorphic to
a right-shifted down-set.
\end{theorem}

\begin{proof}
We will show that any optimal set is isomorphic to a
right-shifted set.  The proof that it must be isomorphic to a down-set
as well is similar.  A similar proof for distance-sum optimal sets
(see Section~\ref{subsec:large_garble}) was given by K\"{u}ndgen in
\cite{kundgen}.

Recall that
$$\sP_\cS(\ga) = \frac{(1-\ga)^n}{|\cS|}
\sum_{\xx,\yy \in \cS} \ze^{d(\xx,\yy)},$$
where $\ze = \ga/(1-\ga) \in (0,1)$.  If $\cS$ is not right-shifted,
there is some $\xx\in \cS$ with $x_i=1$, $x_j=0$, and $i<j$.
Let $\varphi_{ij}(\cS)$ replace all such sets $\xx$ with $s_{ij}(\xx)$.
We only need to
show that this will not decrease
$\sP_\cS(\ga)$.

Consider such an $\xx$ and any $\yy \in \cS$.  If $y_i = y_j$, then
$d(\xx,\yy) = d(s_{ij}(\xx),\yy)$, and $P_\cS(\ga)$ will not change.
If $y_i = 0$  and $y_j=1$, then
$d(\xx,\yy) = d(s_{ij}(\xx),\yy)-2$, and since $\ze^{l-2} \geq \ze^l$,
that term's contribution to $P_\cS(\ga)$ increases.

Suppose $y_i = 1$  and $y_j=0$.  If $s_{ij}(\yy) \in \cS$, then
$d(\xx,\yy)+d(\xx,s_{ij}(\yy))=d(s_{ij}(\xx),\yy)+d(s_{ij}(\xx),s_{ij}(\yy))$,
and $P_\cS(\ga)$ is unchanged.  Otherwise, $\varphi_{ij}(\cS)$ will
replace $\yy$ by $s_{ij}(\yy)$, and $d(\xx,\yy) =
d(s_{ij}(\xx),s_{ij}(\yy))$ means that $P_\cS(\ga)$ will again be unchanged.
\end{proof}


Let $R_{s,n}$ denote an optimal set of size $s$ in $\Fnum_2^n$.
By computing all right-shifted down-sets of size $2^t$, for $t \leq 6$,
we have the following result:

\begin{theorem}
\label{thm:nauty_results}
The optimal sets $R_{2^t,n}$ for $t \in \{1,\dots,6\}$
correspond to Tables~\ref{tab:opt} [pg.~\pageref{tab:opt}] and
\ref{tab:opt64}
[pg.~\pageref{tab:opt64}].
\end{theorem}

These figures, and details of the computations, are given the
Appendix.  Some of the optimal sets for $t=6$ do better than the
sets corresponding to the codes in Figure~\ref{fig:cross}.

\subsection{Optimal Sets for Large Error Probabilities} 
\label{subsec:large_garble}

Theorem~\ref{thm:gamma_small}
states that for any $n$ and
$k$,
for a sufficiently small
error probability $\ga$, a $2^{n-k}$-subcube is an optimal set.
One may also ask what an optimal set is
at the other extreme, a large error probability.
In this section we use existing results about
minimum average distance subsets to list additional sets
that are optimal as $\ga \to 1/2^-$.

We have
\begin{eqnarray*}
 \sP_\cS(\ga) & := &
 \frac{(1-\ga)^n}{|\cS|} A\left(\cS ,\frac{\ga}{1-\ga} \right) \\
 &=& \frac{1}{|\cS|} \sumno_i A_i \ga^i (1-\ga)^{n-i} \, .
\end{eqnarray*}
Letting $\ga = 1/2-\ep$ and $s = |\cS|$, $\sP_\cS(\gamma)$ becomes
\begin{eqnarray*}
\lefteqn{s^{-1} \sumno_i A_i \left( 1/2 - \ep \right)^i
   \left( 1/2 + \ep \right)^{n-i}}
  \qquad &&  \\
 &=&
  \frac{1}{s \, 2^n} \left(
     \sumno_i A_i + \ep \left(  \sumno_i 2 (n-2i) A_i \right) + O(\ep^2)
    \right) \\
 &=&
  \frac{s}{2^n} ( 1 + 2 n \ep ) - \frac{4 \ep}{s \, 2^n} \sumno_i i A_i
   + O(\ep^2 ) \, .
\end{eqnarray*}
Therefore, an optimal set for $\ga \to 1/2^-$ must minimize
the \emph{distance-sum}
of $\cS$
\begin{eqnarray}
 d(\cS)
 &:=& \frac{1}{2} \sum_{\xx,\yy \in \cS} d(\xx,\yy)
 = \frac{1}{2} \sumno_i i A_i \, .
\end{eqnarray}


Denote the minimal distance sum by
\begin{eqnarray*}
 f(s,n) := \min \left\{
  d(\cS): \cS \subset \Fnum_2^n, \, |\cS|=s
 \right\} \, .
\end{eqnarray*}
If $d(\cS) = f(s,n)$ for a set $\cS$ of size $s$,
we say that $\cS$ is \emph{distance-sum optimal}.
The question of which sets are distance-sum optimal was proposed by
Ahlswede and Katona in 1977; see K\"{u}ndgen \cite{kundgen} for
references and recent results.

This question is also difficult.  K\"{u}ndgen presents distance-sum
optimal sets for small $s$ and $n$, which include the ones of size
16 from Table~\ref{tab:opt}.  Jaeger et al. \cite{jaeger} found the
distance-sum optimal set for $n$ large.

\begin{theorem}
{}
(Jaeger, et al.~\cite{jaeger}, cf.~\cite[pg.~151]{kundgen})
  For $n \geq s-1$, a generalized 1-sphere (with $s$ points)
is distance-sum optimal
unless $s \in \{ 4, 8 \}$ (in which case the subcube is optimal).
\end{theorem}

From this we have:

\begin{corollary}
  For $n \geq 2^t-1$, with $t\geq 4$ and $\ga$ sufficiently close to $1/2$, a
  $(2^t-1)$-dimensional 1-sphere is hashing optimal.
\end{corollary}

\section{Hashes from Random Codes}
\label{sec:shannon}

In this section we will show that hashes from
random
codes under minimum weight
decoding\footnote{Ties arising in minimum weight decoding are
broken in some unspecified manner.}
perform better than projection.
Let $R=k/n$ be the rate of a code.
The error exponent
for $k$-projection, $E^{\cP_{n,k}}(p)$, is
\begin{equation}
  \label{eq:proj_asymp}
 -\frac{1}{n} \lg {\cP^{n,k}}(p) = -\frac{1}{n} \lg (1-p)^k = -R \lg (1-p).  
\end{equation}

Theorem~\ref{thm:hamming_wins} shows that for any $\ga>0$ there are
codes with rate $R \approx 1$ which beat projection.
For any fixed $R$, we will bound
the expected error exponent for
a random
code $\cR$ of rate $R$,
and show that it beats (\ref{eq:proj_asymp}).

Let $H$ be the binary entropy
\begin{equation}
\label{eq:entropy}
H(\de) := - \de \lg \de - (1-\de) \lg (1-\de) \, .
\end{equation}

Fix $\delta \in [0,1/2)$.
Let $d := \lfloor \delta n \rfloor$,
let $\cS_d(\xx)$ denote the sphere of
radius $d$ around $\xx$, and let $V(d) := |\cS_d(\xx)|$.

It is elementary to show (see \cite{gallager}, Exercise 5.9):
\begin{lemma}\label{lem:gv}
  Let $\cR$ be a random
 code of length $n$ and rate $R$, where
  $n$ is sufficiently large.  For
  $\cc\in \cR$, the probability that a given vector $\xx \in
  \cS_d(\cc)$ is closer to another codeword than $\cc$ is at most
$$
2^{n (H(\delta)-1+R)}.    
$$
\end{lemma}
Lemma~\ref{lem:gv} implies that if 
$H(\de) < 1-R$
(the Gilbert-Varshamov bound), then with high probability, any given $\xx \in
\cS_d(\cc)$ will be decoded to $\cc$.
For the rest of this section we will assume this bound, 
so that Lemma~\ref{lem:gv} applies.

Let $\sP^\cR(\ga)$ be the probability that a random point $\xx$
and $\xx+\gg$ both hash to $\cc$.  This is greater than
the probability that
$\xx+\gg$ has weight exactly $d$, so
$$
\sP^\cR(p) > \sum_{i=0}^d {d \choose i} {{n-d} \choose i} p^{2i} (1-p)^{n-2i}.
$$
Theorem 4 of \cite{ackl} gives a bound for this:

\begin{theorem}\label{thm:emax}
For any $\epsi \leq 1/2$ and $\de$ such that $H(\de)<1-R$ and $\epsi
\leq 2\de$,
\begin{eqnarray*}
-\sE^\cR(\ga)
&\geq & \epsi \lg \ga + (1-\epsi) \lg (1-\ga) \\
& + &  \de H \left( \frac{\epsi}{2\de} \right)
 + (1-\de) H \left( \frac{\epsi}{2(1-\de)} \right) 
\end{eqnarray*}
for any $\epsi \leq 1/2$.  The right hand side is maximized at
$\epsimax$ satisfying
\begin{eqnarray*}
 \frac{(2\de-\epsimax)(2(1-\de)-\epsimax)}{\epsimax^2}
  &=& \frac{(1-\ga)^2}{\ga^2} .
\end{eqnarray*}
\end{theorem}

Define
\begin{eqnarray*}
D(\ga,\delta,\epsi) &:=& \epsi \lg \ga + (1-\epsi) \lg (1-\ga)
 + \de H \left( \frac{\epsi}{2\de} \right)\\
&& \hbox{}  + (1-\de) H \left( \frac{\epsi}{2(1-\de)} \right) \\
&& \hbox{}   -(1-H(\de)) \lg(1-p)  \, .
\end{eqnarray*}
Then $E^{\cP_{n,k}}(p)-\sE^\cR(\ga) \geq D(\ga,\delta,\epsi)$.

\begin{theorem}\label{thm:Dpos}
$D(\ga,\de,\epsimax) > 0$ for any $\de, \ga \in (0,1/2)$.
\end{theorem}

\begin{proof}
Fix $\de \in (0,1/2)$, and let $f(\ga) := D(\ga,\de,\epsimax)$.  It is
easy to check that:
\begin{eqnarray*}
&&  \lim_{\ga \rightarrow 0^+} f(\ga)  =  0, \\
&&  \lim_{\ga \rightarrow 1/2^-} f(\ga)  =  0, \\
&&  \lim_{\ga \rightarrow 0^+} f'(\ga)  >  0, \\
&&  \lim_{\ga \rightarrow 1/2^-} f'(\ga)  <  0, \\
\end{eqnarray*}
Therefore, it suffices to show that $f'(\ga)$ has only one zero in
$(0,1/2)$.
Observe that $\epsimax$ is chosen so that
$\frac{\partial D}{\partial \epsi}(\de,\ga,\epsimax) = 0.$
Hence
\begin{eqnarray*}
  f'(p) &=& \frac{\partial D}{\partial \ga}(\de,\ga,\epsimax)\\
&=& \frac{\epsimax}{\ga \log(2)}
- \frac{1-\epsimax}{(1-\ga) \lg(2)}
+ \frac{1-H(\de)}{(1-\ga) \log(2)},
\end{eqnarray*}
so
\begin{eqnarray*}
\log(2) f'(\ga) &=&
 \frac{\epsimax}{\ga} - \frac{1-\epsimax}{1-\ga} + \frac{1-H(\de)}{1-\ga}.
\end{eqnarray*}
Therefore $f'(\ga)=0$ when $\epsimax = \ga H(\de)$.
From Theorem~\ref{thm:emax}
we find
$$
\ga = \frac{ 4\de(1-\de) - H(\de)^2 }{2  (H(\de) - H(\de)^2) }.
$$
\end{proof}

Thus we have 
$E^{\cP_{n,k}}(p) > \sE^\cR(\ga)$, and so:

\begin{corollary}\label{thm:codes_win}
For any $\ga \in (0, 1/2)$,
$R \in (0,1)$ and $n$ sufficiently large,
the expected probability of collision for
a random code of rate $R$ is higher than
projection.
\end{corollary}

\section*{Acknowledgements.}
The authors would like to thank William Bradley, David desJardins and
David Moulton for stimulating discussions which helped initiate this
work.  Also, Tom Dorsey and Amit Khetan provided the simpler proof
of Theorem~\ref{thm:Dpos} given here.  The anonymous referees made a
number of good suggestions that improved the paper, particularly the
exposition in the introduction.


\appendix
\label{sec:details}

By Theorem~\ref{thm:rightdown}, we may find all optimal sets by
examining all right-shifted down-sets.  Right-shifted down-sets
correspond to ideals in the poset whose elements are in $\Fnum_2^n$
and with partial order $\xx \preceq \yy$ if $\xx$ can be obtained from
$\yy$ by a series of $\rho_i$ (\ref{eq:rho})
and $\si_{ij}$ (\ref{eq:si}) operations.  It turns out
that there are not too many such ideals, and they may be computed
efficiently.

Our method for producing the ideals is not new, but
since the main references are unpublished, we describe them briefly
here.  In Section 4.12.2 of \cite{ruskey}, Ruskey
describes a procedure GenIdeal for listing the ideals in a poset
$\cP$.
Let $\ddd{\xx}$ denote all the elements $\preceq \xx$, and
$\uuu{\xx}$ denote all the elements $\succeq \xx$.

\begin{list}{\hspace*{3em}}{\renewcommand{\parsep}{0pt}}
\item {\bf procedure} GenIdeal($\cQ$: Poset,$\,$ $I$: Ideal)
\item {\bf local} $\xx$: PosetElement
\item {\bf begin}
\item \hspace*{2em} {\bf if} $\cQ = \phi$ {\bf then} PrintIt($I$);
\item \hspace*{2em} {\bf else}
\item \hspace*{4em} $\xx := $ some element in $\cQ$;
\item \hspace*{4em} GenIdeal($\cQ - \ddd{\xx}, \; I \cup \ddd{\xx})$;
\item \hspace*{4em} GenIdeal($\cQ - \uuu{\xx}, \; I)$;
\item {\bf end}
\end{list}

The idea is to start with $I$ empty, and $\cQ= \cP$.  Then
for each $\xx$, an ideal either contains $\xx$, in which case it will be
found by the first call to GenIdeal, or it does not, in which case the
second call will find it.

Finding $\uuu{\xx}$ and $\ddd{\xx}$ may be done efficiently if we precompute
two $|\cP| \times |\cP|$ incidence matrices representing these sets
for each element of $\cP$.  This precomputation takes time
$O(|\cP|^2)$, and then the time per ideal is $O(|\cP|)$.
This is independent of
the choice of $\xx$.  Squire (see \cite{ruskey} for details)
realized that, by picking $\xx$ to be the
middle element of $\cQ$ in some linear extension, the time per ideal
can be shown to be $O(\lg |\cP|)$.

We are only interested in down-sets that are right-shifted and also
are of fairly small size.  The feasibility of our computations
involves both issues.  In particular, within GenIdeal we may restrict
to $\xx \in \Fnum_2^n$ with $\Size(\ddd{\xx})$ no more than the target
size of the set we are looking for.  If we were using GenIdeal with the
poset whose ideals correspond to down-sets
of size 64 in $\Fnum_2^{63}$, there would be
$83,278,001$ such $\xx$ to consider.
However, for our situation with right-shifted down-sets,
there are only 257 such $\xx$
and the problem becomes quite manageable.
Furthermore,
instead of stopping when $\cQ$ is empty, we
stop when $I$ is at or above the desired size.

Table~\ref{tab:rightdown} gives the number of right-shifted down-sets
of different sizes.
The computation for size 32 sets took just over a second on one processor of
an HP Superdome.  Size 64 sets took 23 minutes.
Let $R_{s,n}$ refer to an optimal set of size $s$ in $\Fnum_2^n$.
Tables~\ref{tab:opt} and \ref{tab:opt64} list
$R_{2^t,n}$ for all $t\leq 6$ and all $n < 2^t$.

\begin{table}[!t]
\caption{Number of right-shifted down-sets}
\label{tab:rightdown}
\small
\centering
\begin{tabular}{c@{\quad\quad}c@{\quad\quad}c}
\begin{tabular}{|c|c|}
\hline
size & number  \\
\hline\hline
2 & 1 \\
3 & 1 \\
4 & 2 \\
5 & 2 \\
6 & 3 \\
7 & 4 \\
8 & 6 \\
9 & 7 \\
10 & 10 \\
\hline
\end{tabular}
&
\begin{tabular}{|c|c|}
\hline
size & number  \\
\hline\hline
11 & 13 \\
12 & 18 \\
13 & 23 \\
14 & 31 \\
15 & 40 \\
16 & 54 \\
17 & 69 \\
18 & 91 \\
19 & 118 \\
20 & 155 \\
\hline
\end{tabular}
&
\begin{tabular}{|c|c|}
\hline
size & number  \\
\hline\hline
21 & 199 \\
22 & 260 \\
23 & 334 \\
24 & 433 \\
32 & 3140 \\
48 & 130979 \\
64 & 4384627  \\
\hline
\end{tabular}
\end{tabular}
\end{table}

\begin{table}[!t]
  \caption{Optimal right-shifted down-sets $R_{64,n}$ beating known codes.
 (There are no such down-sets $R_{2^t,n}$ for $t \leq 5$.)}
  \label{tab:optk}
  \centering
  \begin{tabular}{|c||c|l|l|}
\hline
$k$ & $n$ & cross & $R_{64,n}$ \\
\hline\hline
6 & 12 & $0.487$ & $\paira{2^{11},2^{10}+2^5,3\cdot 2^8}$ \\
7 & 13 & $0.470$ &  $\paira{2^{12},2^{10}+2^4,3\cdot 2^8}$ \\
8 & 14 & $0.439$ & $\paira{2^{13}+2^2,2^{13}+3,2^3+2^2+1}$ \\
9 & 15 & $0.391$ & $\paira{2^{14}+3,2^{10}+2^2}$ \\
16 & 22 & $0.244$ & $\paira{2^{21}+2}$ \\
17 & 23 & $0.242$ & $\paira{2^{22}+1,2^{19}+2}$ \\
18 & 24 & $0.238$ & $\paira{2^{23}+1,2^{17}+2}$ \\
19 & 25 & $0.231$ & $\paira{2^{24}+1,2^{15}+2}$ \\
20 & 26 & $0.222$ & $\paira{2^{25}+1,2^{13}+2}$ \\
21 & 27 & $0.212$ & $\paira{2^{26}+1,2^{11}+2}$\\
\hline
  \end{tabular}
\end{table}

\begin{table*}[!t]
  \caption{Optimal right-shifted down-sets $R_{2^t,n}$ ($t \leq 5$).}
  \label{tab:opt}
  \centering
\small
  \begin{tabular}{|c|c||c|l|l|}
\hline
$t$ & $n$ & $\ga_{\rm cross}$ & distance distribution function & $R_{2^t,n}$ \\
\hline\hline
1 &  1 & $0$ & $2(1+x)$  & $\paira{1}$ \\
\hline
2 &  2 & $0$ & $4(1+x)^2$  & $\paira{2^2-1}$ \\
\hline
3 &  3 & $0$ & $8(1+x)^3$  & $\paira{2^3-1}$ \\
\hline\hline
4 &  4     & $0$ & $16(1+x)^4$  & $\paira{2^4-1}$ \\
  & 12 & $0.4560$ & $16+36x+144x^2+60x^3$  & $\paira{2^{11},2^{3}+1}$ \\
  & '' & ''       & \qquad\qquad '' & $\paira{2^{11},3\cdot2}$\\
  & 13 & $0.3929$ & $16+34x+162x^2+44x^3$  & $\paira{2^{12},2^2+1}$ \\
  & 14 & $0.3333$ & $16+32x+184x^2+24x^3$  & $\paira{2^{13},2+1}$\\
  & 15 & $0.2826$ & $16+30x+210x^2$ & $\paira{2^{14}}$ \\
\hline\hline
5 &  5 & $0$        & $32(1+x)^5$  &  $\paira{2^5-1}$ \\
  & 12 & $0.4882$         & $32+100x +368x^2+ 380x^3+144x^4$  & $\paira{2^{11}+1,2^9+2}$ \\
  & '' & '' & \qquad\qquad '' & $\paira{2^{11},2^{10}+2}$ \\
  & 13 & $0.4492$         & $32+98x +378x^2+ 396x^3+120x^4$  & $\paira{2^{12}+1,2^7+2}$\\
  & 14 & $0.3929$         & $2(1+x)(16+34x+162x^2+44x^3)$ &  $\paira{2^{13}+1,2^3+3}$ \\
  & 15 &$0.3333$          & $2(1+x)(16+32x+184x^2+24x^3)$ & $\paira{2^{14}+1,7}$ \\
  & 16 & $0.2826$   & $2(1+x)(16+30x+210x^2)$ & $\paira{2^{15}+1}$ \\
  & \fbox{19} & $0.3333$        & $32+86x +498x^2+ 408x^3$  & $\paira{2^{18},2^{12}+1}$ \\
  & 20 & $0.2799$         & $32+84x +512x^2+ 396x^3$  & $\paira{2^{19},2^{11}+1}$\\
  & 21 & $0.2724$         & $32+82x +530x^2+ 380x^3$  & $\paira{2^{20},2^{10}+1}$\\
  & 22 & $0.2627$         & $32+80x +552x^2+ 360x^3$  & $\paira{2^{21},2^{9}+1}$\\
  & 23 & $0.2515$         & $32+78x +578x^2+ 336x^3$  & $\paira{2^{22},2^{8}+1}$\\
  & 24 & $0.2390$         & $32+76x +608x^2+ 308x^3$  & $\paira{2^{23},2^{7}+1}$\\
  & 25 & $0.2259$         & $32+74x +642x^2+ 276x^3$  & $\paira{2^{24},2^{6}+1}$\\
  & 26 & $0.2126$         & $32+72x +680x^2+ 240x^3$  & $\paira{2^{25},2^{5}+1}$\\
  & 27 & $0.1992$         & $32+70x +722x^2+ 200x^3$  & $\paira{2^{26},2^{4}+1}$\\
  & 28 & $0.1864$         & $32+68x +768x^2+ 156x^3$  & $\paira{2^{27},2^{3}+1}$ \\
  & '' & '' & \qquad\qquad '' & $\paira{2^{27},3\cdot 2}$ \\
  & 29 & $0.1741$   & $32+66x +818x^2+ 108x^3$  & $\paira{2^{28},2^{2}+1}$\\
  & 30 & $0.1626$   & $32+64x +872x^2+ 56x^3$  & $\paira{2^{29},2+1}$\\
  & 31 & $0.1518$ & $32+62x +930x^2$ & $\paira{2^{30}}$ \\
\hline
  \end{tabular}
\end{table*}

Several features of Tables~\ref{tab:opt} and \ref{tab:opt64}
require explanation.
First we identify the binary expansion $x = \sum_{i<n} 2^i x_{n-i}$ with
the vector $\xx = (x_1,\dots,x_n)$.
Second, for each optimal right-shifted down-set $R_{2^t,n}$
we have listed a minimal set of generators.
For example $\paira{2^4-1}$ corresponds to the $4$-dimensional cube
while $\paira{2^{14}}$, as a subset of $\Fnum_2^{15}$, corresponds
to the $15$-dimensional 1-sphere.

For each set $\ga_{\rm cross}$ indicates the crossover value for
$\ga$ at which point that set performs better than any preceding
entry in the table.  For example, the 4-dimensional cube
$\paira{2^4-1}$ is optimal for all $\ga \in (0,0.5)$ if $4 \leq n \leq
11$ but is only optimal for $\ga \in (0,0.4560)$ if $n=12$.
For $(t,n)=(4,13)$, the 4-dimensional cube
is optimal for $\ga \in (0,\,0.3929)$
while the right-shifted down-set $\paira{2^{12},2^2+1}$
is optimal for $\ga \in (0.3929, \, 0.5)$.

There are several specific $(t,n)$ for which more than two
nonisomorphic right-shifted down-sets are optimal.  In several cases
the nonisomorphic optimal right-shifted down-sets have the same
distance distribution.  (The two nonisomorphic sets $R_{2^4,12}$
were originally found by K\"{u}ndgen \cite[pg.~160:
Table~1]{kundgen}.)  In other cases different sets are optimal for
different values of $\ga$.  (Such cases are highlighted
with a box \fbox{$\cdot$}.)
For example, with $(t,n)=(5,19)$, the
5-dimensional cube $\paira{2^5-1}$ is optimal for $\ga \in (0, \,
0.2826)$, $\paira{2^{15}+1}$ is optimal on $(0.2826, \, 0.3333)$,
while $\paira{2^{18},2^{12}+1}$ is optimal on $(0.3333, \, 0.5)$.
Somewhat similar situations involve $t=6$ and $n \in \{19, 28, 29, 35,
36, 37, 38, 58, 59\}$.\footnote{For $(t,n)=(6,28)$, the three sets are
$\paira{2^6-1}$ on $(0,0.199)$, $\paira{2^{27}+1,2^5+3}$ on
$(0.199,0.25)$ and $\paira{2^{27}+1,2^{9}+2}$ on $(0.25,0.5)$.}  For
$t \leq 6$ and for any $n$, there are at most three different optimal
sets.

\begin{table*}[htbp]
  \caption{Optimal right-shifted down-sets $R_{64,n}$ ($t=6$)}
  \label{tab:opt64}
  \centering
  \tiny
  \begin{tabular}{|c||c|l|l|}
\hline
$n$ & $\ga_{\rm cross}$ & distance distribution function & $R_{64,n}$ \\
\hline\hline
6 & $0$ & $64 + 384 x + 960 x^2 + 1280 x^3 + 960 x^4 + 384 x + 64$ & $\paira{2^6-1}$ \\
12 & $0.487$ & $64 + 228 x + 1092 x^2 + 1020 x^3 + 1692 x^4$ & $\paira{2^{11},2^{10}+2^5,3\cdot 2^8}$ \\
13 & $0.470$ & $64 + 226 x + 1086 x^2 + 1100 x^3 + 1620 x^4$ & $\paira{2^{12},2^{10}+2^4,3\cdot 2^8}$ \\
14 & $0.439$ & $64 + 250 x + 1002 x^2 + 1508 x^3 + 1032 x^4 + 240 x^5$ & $\paira{2^{13}+2^2,2^{13}+3,2^3+5}$ \\
15 & $0.391$ & $64 + 248 x + 1024 x^2 + 1592 x^3 + 992 x^4 + 176 x^5$ & $\paira{2^{14}+3,2^{10}+2^2}$ \\
16 & $0.333$ & $4(1 + x)^2 (16 + 32 x + 184 x^2 + 24 x^3)$ & $\paira{2^{15}+3,2^4-1}$ \\
17 & $0.283$ & $4(1+x)^2 (16+30x+210x^2)$ & $\paira{2^{16}+3}$ \\
\fbox{19} & $0.36$ & $64 + 232 x + 1184 x^2 + 1784 x^3 + 832 x^4$ & $\paira{2^{18}+2,2^{10}+3}$\\
20 & $0.277$ & $64 + 224 x + 1240 x^2 + 1752 x^3 + 816 x^4$ & $\paira{2^{19}+2,2^{7}+3}$\\
21 & $0.263$ & $64 + 216 x + 1320 x^2 + 1704 x^3 + 792 x^4$ & $\paira{2^{20}+2,2^{4}+3}$\\
22 & $0.244$ & $64 + 208 x + 1424 x^2 + 1640 x^3 + 760 x^4$ & $\paira{2^{21}+2}$ \\
23 & $0.242$ & $64 + 206 x + 1426 x^2 + 1680 x^3 + 720 x^4$ & $\paira{2^{22}+1,2^{19}+2}$ \\
24 & $0.238$ & $64 + 204 x + 1440 x^2 + 1716 x^3 + 672 x^4$ & $\paira{2^{23}+1,2^{17}+2}$ \\
25 & $0.231$ & $64 + 202 x + 1466 x^2 + 1748 x^3 + 616 x^4$ & $\paira{2^{24}+1,2^{15}+2}$ \\
26 & $0.222$ & $64 + 200 x + 1504 x^2 + 1776 x^3 + 552 x^4$ & $\paira{2^{25}+1,2^{13}+2}$ \\
27 & $0.212$ & $64 + 198 x + 1554 x^2 + 1800 x^3 + 480 x^4$ & $\paira{2^{26}+1,2^{11}+2}$\\
\fbox{28} & $0.199$ & $2(1+x)(32+70x+722x^2+200x^3)$ & $\paira{2^{27}+1,2^5+3}$ \\
'' & $0.25$ & $64 + 196 x + 1616 x^2 + 1820 x^3 + 400 x^4$ & $\paira{2^{27}+1,2^{9}+2}$\\
\fbox{29} & $0.186$ & $2(1+x)(32+68x+768x^2+156x^3)$ & $\paira{2^{28}+1,2^4+3}$ \\
 '' & '' & \qquad\qquad '' &  $\paira{2^{28}+1,3\cdot 2^2 +1}$ \\
'' & $0.333$ & $64 + 194 x + 1690 x^2 + 1836 x^3 + 312 x^4$ & $\paira{2^{28}+1,2^7+2}$\\
30 & $0.174$ & $2(1+x)(32+66x+818x^2+108x^3)$ & $\paira{2^{29}+1,2^3+3}$ \\
31 & $0.163$ & $2(1+x)(32+64x+872x^2+56x^3)$ & $\paira{2^{30}+1,7}$\\
32 & $0.152$ & $2(1+x)(32+62x+930x^2)$ & $\paira{2^{31}+1}$\\
\fbox{35} & $0.1538$ & $64 + 182 x + 2002 x^2 + 1848 x^3$ & $\paira{2^{34},2^{28}+1}$\\
\fbox{36} & $0.1537$ & $64 + 180 x + 2016 x^2 + 1836 x^3$ & $\paira{2^{35},2^{27}+1}$\\
\fbox{37} & $0.153$ & $64 + 178 x + 2034 x^2 + 1820 x^3$ & $\paira{2^{36},2^{26}+1}$\\
\fbox{38} & $0.152$ & $64 + 176 x + 2056 x^2 + 1800 x^3$ & $\paira{2^{37},2^{25}+1}$\\
39 & $0.151$ & $64 + 174 x + 2082 x^2 + 1776 x^3$ & $\paira{2^{38},2^{24}+1}$\\
40 & $0.150$ & $64 + 172 x + 2112 x^2 + 1748 x^3$ & $\paira{2^{39},2^{23}+1}$\\
41 & $0.148$ & $64 + 170 x + 2146 x^2 + 1716 x^3$ & $\paira{2^{40},2^{22}+1}$\\
42 & $0.146$ & $64 + 168 x + 2184 x^2 + 1680 x^3$ & $\paira{2^{41},2^{21}+1}$\\
43 & $0.144$ & $64 + 166 x + 2226 x^2 + 1640 x^3$ & $\paira{2^{42},2^{20}+1}$\\
44 & $0.141$ & $64 + 164 x + 2272 x^2 + 1596 x^3$ & $\paira{2^{43},2^{19}+1}$\\
45 & $0.139$ & $64 + 162 x + 2322 x^2 + 1548 x^3$ & $\paira{2^{44},2^{18}+1}$\\
46 & $0.136$ & $64 + 160 x + 2376 x^2 + 1496 x^3$ & $\paira{2^{45},2^{17}+1}$\\
47 & $0.133$ & $64 + 158 x + 2434 x^2 + 1440 x^3$ & $\paira{2^{46},2^{16}+1}$\\
48 & $0.130$ & $64 + 156 x + 2496 x^2 + 1380 x^3$ & $\paira{2^{47},2^{15}+1}$\\
49 & $0.127$ & $64 + 154 x + 2562 x^2 + 1316 x^3$ & $\paira{2^{48},2^{14}+1}$\\
50 & $0.123$ & $64 + 152 x + 2632 x^2 + 1248 x^3$ & $\paira{2^{49},2^{13}+1}$\\
51 & $0.120$ & $64 + 150 x + 2706 x^2 + 1176 x^3$ & $\paira{2^{50},2^{12}+1}$\\
52 & $0.117$ & $64 + 148 x + 2784 x^2 + 1100 x^3$ & $\paira{2^{51},2^{11}+1}$\\
53 & $0.114$ & $64 + 146 x + 2866 x^2 + 1020 x^3$ & $\paira{2^{52},2^{10}+1}$\\
54 & $0.110$ & $64 + 144 x + 2952 x^2 + 936 x^3$ & $\paira{2^{53},2^{9}+1}$\\
55 & $0.107$ & $64 + 142 x + 3042 x^2 + 848 x^3$ & $\paira{2^{54},2^{8}+1}$\\
56 & $0.104$ & $64 + 140 x + 3136 x^2 + 756 x^3$ & $\paira{2^{55},2^7+1}$\\
57 & $0.101$ & $64 + 138 x + 3234 x^2 + 660 x^3$ & $\paira{2^{56},2^6+1}$\\
\fbox{58} & $0.0978$ & $64 + 138 x + 3330 x^2 + 452 x^3 + 112 x^4$ & $\paira{2^{57},2^3+1,3\cdot 2}$ \\
'' & $0.1047$ & $64 + 136 x + 3336 x^2 + 560 x^3$ & $\paira{2^{57},2^5+1}$ \\
\fbox{59} & $0.0946$ & $64 + 136 x + 3440 x^2 + 344 x^3 + 112 x^4$ & $\paira{2^{58},7}$ \\
'' & $0.1179$ & $64 + 134 x + 3442 x^2 + 456 x^3$ & $\paira{2^{59},2^4+1}$\\
60 & $0.0920$ & $64 + 132 x + 3552 x^2 + 348 x^3$ & $\paira{2^{59},2^3+1}$ \\
'' & '' & \qquad\qquad '' & $\paira{2^{59},3\cdot 2}$ \\
61 & $0.0891$ & $64 + 130 x + 3666 x^2 + 236 x^3$ & $\paira{2^{60},2^2+1}$\\
62 & $0.0864$ & $64 + 128 x + 3784 x^2 + 120 x^3$ & $\paira{2^{61},2+1}$ \\
63 & $0.0838$ & $64 + 126 x + 3906 x^2$ & $\paira{2^{62}}$\\
\hline
  \end{tabular}
\end{table*}

Some of the optimal sets $R_{64,n}$ are better than those for any known hash
function.  Table~\ref{tab:optk} gives the best known sets for each
$k$, and their generators.

Tilings of binary spaces have also been studied \cite{tilings}.  Indeed
a complete translation-invariant decoding algorithm leads to a tiling
of the $n$-cube.
Recently the second author and Coppersmith \cite{cm} have shown that
none of these optimal sets are associated to tilings.





%





\end{document}